\documentclass{webofc}
\usepackage[varg]{txfonts}   % Web of Conferences font
\begin{document}
\title{Can we observe effects of early nonequilibrium dynamics using collective flow}

\author{\firstname{Piotr} \lastname{Bo{\.z}ek}\inst{1}\fnsep\thanks{\email{piotr.bozek@fis.agh.edu.pl}}
        % etc.
}

\institute{AGH University of Krakow, Faculty of Physics and Applied Computer Science, aleja Mickiewicza 30, 30-059 Cracow, Poland
          }

\abstract{%
  The rapid expansion of the fireball created in a heavy-ion collision causes strong departures from local equilibrium. Such effects are especially important in the very early phase of the collision, bringing a substantial pressure asymmetry. We investigate effects of this early stage pressure asymmetry in a kinetic model without boost-invariance. In the kinetic evolution the pressure asymmetry results from the competition of the longitudinal expansion and the kinetic relaxation. Unlike the transverse flow, the directed flow is found to be very sensitive to the equilibration rate and the pressure anisotropy. 
}
\maketitle
\section{Introduction}
\label{intro}

One of the main goals of the experimental program in
ultrarelativistic heavy-ion collisions at the BNL Relativistic Heavy Ion Collider and the CERN Large Hadron Collider is to study the properties of the quark-gluon plasma formed in the collision \cite{Heinz:2013th,Gale:2013da}. The dense matter formed in the interaction region of an ultrarelativistic heavy-ion collision expands very fast. The equilibrium properties of the deconfined state of matter can be simulated in lattice QCD simulations. On the other hand, the strong expansion of the fireball in heavy-ion collisions generates
large deviations from local thermal equilibrium. The nonequilibrium effects are important if the expansion rate dominates over the equilibration rate of the matter. Deviations from equilibrium are especially important in two instances:
at the late stages, when freeze-out occurs, and at the very early stages when the expansion rate is very high. At late stages the nonequilibrium dynamics can be
modeled realistically using a hadronic cascade.

In this paper, we study effects of nonequilibrium dynamics at the early stage. If the deviations from local equilibrium are important, the dynamics of at the very early stage cannot be reliably described using viscous hydrodynamics \cite{Niemi:2014wta}. It is possible to address the far from equilibrium dynamics using kinetic equations. In the framework of the kinetic description, the fast longitudinal expansion in a heavy-ion collision leads to a nonisotropic momentum distribution
\cite{Florkowski:2013lya}. The effective longitudinal pressure is much smaller than the transverse one. It is interesting to ask the question whether this strong deviation from equilibrium can be observed in the final collective flow observables. For collective flow observables build on the transverse momenta of the  emitted particles, e.g. transverse  and  harmonic flow, the answer is negative. 
The effect of the strong pressure asymmetry on the such collective flow observables is very small  \cite{Vredevoogd:2008id}.  On the other hand, we find that  the effect of the pressure asymmetry on the direct flow is important in the kinetic dynamics \cite{Bozek:2022cjj}.

\section{Kinetic dynamics}
\label{sec-1}

The early dynamics in a heavy-ion collision is often described with a kinetic approach (or simply a free-streaming expansion) \cite{Broniowski:2008qk,Kurkela:2018vqr,daSilva:2022xwu,Ambrus:2022koq,Liyanage:2022nua}. At latter stages, when deviations from local equilibrium are smaller, the viscous hydrodynamic  model is applied. We study the kinetic dynamics at only the very early stage of the collision. The goal is to find observable effects of this early nonequilibrium  dynamics. The full solution of the kinetic equation is technically difficult. The evolution equation
\begin{equation}
  p^\mu\partial_\mu f = -\frac{u^\mu p_\mu}{\tau_R}\left( f-f_{iso}\right) \ 
  \label{eq:rta}
\end{equation} must be solved, for the  multidimensional phase-space distribution $f(t,\vec{x},\vec{p})$.  $f_{iso}$ is an isotropic distribution, e.g. the equilibrium distribution.
A simplification of the equations is possible  for massless particles \cite{Kurkela:2018ygx} using the moment 
\begin{equation}
  F(t,\vec{x},\Omega_p)=\int \frac{p^3 dp}{2 \pi^2} f(t,\vec{x},\vec{p})\ 
  \label{eq:moment}
\end{equation}
of the distribution function. The dimensionality of the phase-space distribution is reduced. $F$ depends only on the spherical angles  $\Omega_p=(\phi,\theta)=(\phi,v_z)$ of the particle  momentum vector $\vec{p}$.
It obeys the following kinetic equation
\begin{equation}
  \partial_t F + \vec{v} \partial_{\vec{x}} F = - \frac{u^\mu v_\mu}{\tau_R}\left( F-F_{iso}\right)  \ .
\end{equation}

\begin{figure}[ht]
\centering
\includegraphics[width=6.2cm,clip]{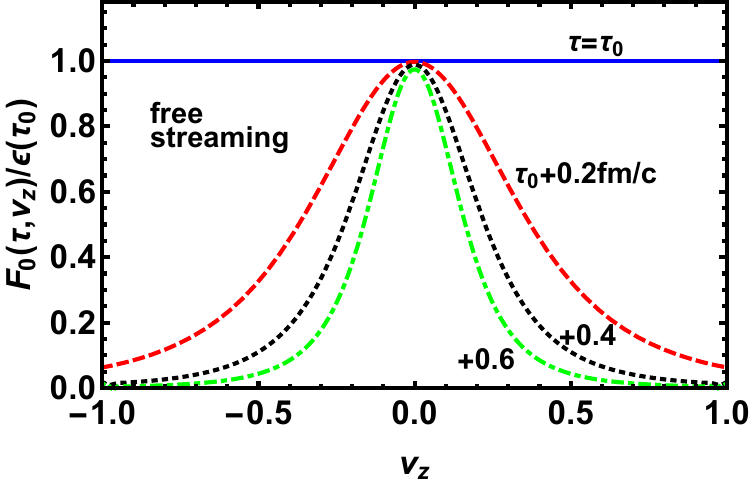}~~~~~\includegraphics[width=6.2cm,clip]{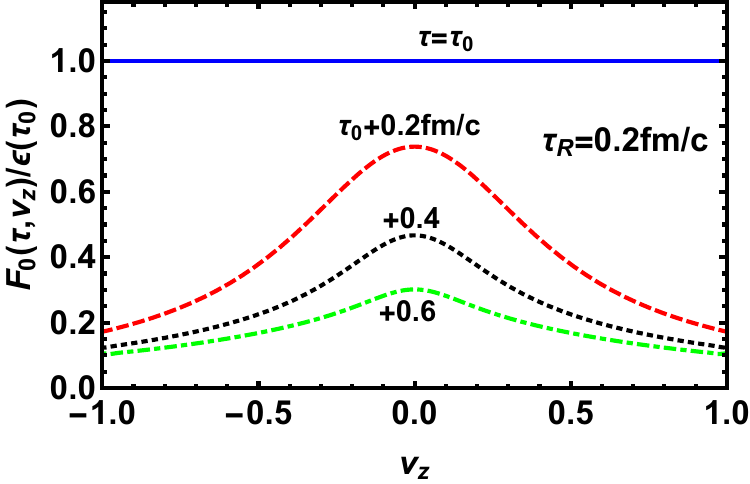}
\caption{The distribution in the longitudinal velocity at different times of the evolution. Left panel: the free-streaming evolution, right panel:  kinetic equation evolution with a finite relaxation time.}
\label{fig:fev}      
\end{figure}

In order to study the development of the rapidity odd directed flow a system without boost invariance must be considered. For simplicity the phase-space distribution is assumed to depend on two space variables, one transverse dimension $x$ and the longitudinal space-time rapidity $\eta$. The initial density is tilted in the $x$-$\eta$ plane away from the beam direction, which generates  a non-zero directed flow \cite{Bozek:2010bi}. 
For a non-boost invariant system the numerical solution is still challenging, with sharply peaked distributions in the longitudinal velocity.
 To reduce the numerical load a rescaled phase-space distribution $\tilde{F}(\tau,x,\eta,\phi,v_z)= F(\tau,x,\eta,\phi,v_z)\left(\cosh(\eta)-v_z\sinh(\eta) \right)^4$ can be used. For details of the solution see \cite{Bozek:2022cjj}. Also the full dependence on the particle longitudinal velocity in the phase-space distribution $F$ is kept in the solution.

We compare three calculations, the free-streaming dynamics and two kinetic equation solutions with two relaxation times. In Fig. \ref{fig:fev} is shown the time evolution of the longitudinal velocity distribution. The narrowing of the distribution is fast for the free-streaming evolution (left panel) and slower when the relaxation term is acting in the kinetic equation (right panel).

\section{Results}

\label{sec-2}

\begin{figure}[ht]
\centering
\includegraphics[width=6.2cm,clip]{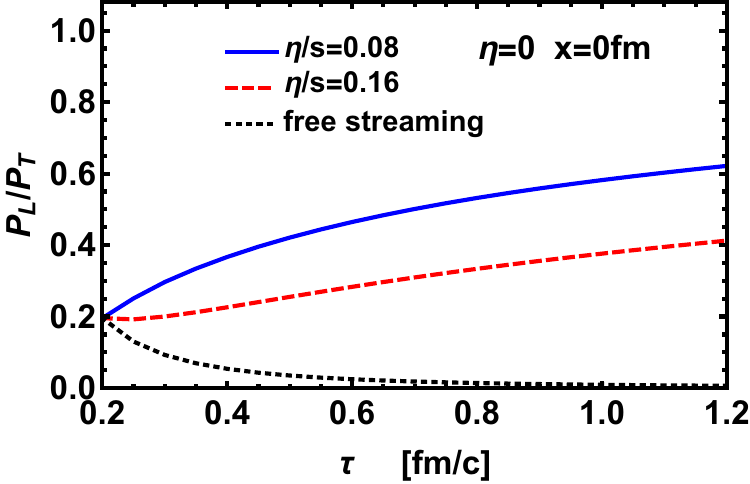}~~~~~\includegraphics[width=6.2cm,clip]{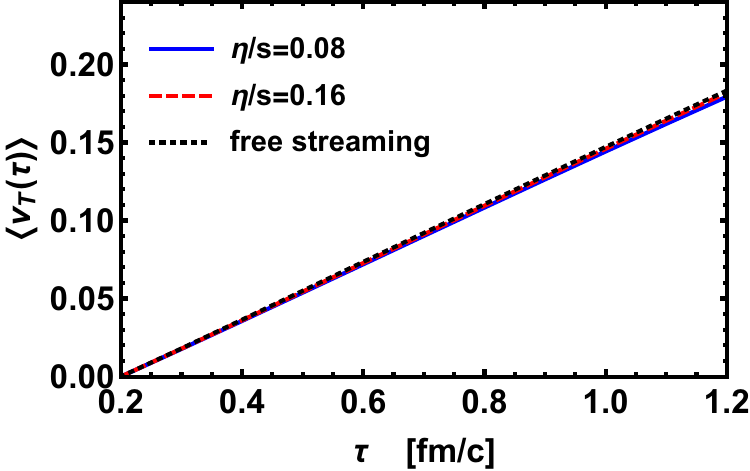}
\caption{(Left panel) The time dependence of the ratio of the longitudinal to transverse pressure at the center of the fireball. (Right panel) The time dependence of the average transverse flow.  Results are  for two values of  the relaxation time and for the free-streaming evolution.  Figs. are from Ref. \cite{Bozek:2022cjj}.}
\label{fig:ptpl}      
\end{figure}

The Bjorken the expansion competes with the isotropization in the kinetic equation. As a result  the pressure in the transverse and longitudinal directions is different (Fig. \ref{fig:ptpl}, left panel). Even at the center of the fireball, where the density is the largest, the pressure never reaches the isotropic equilibrium limit.  The effect of the pressure asymmetry  is dramatic for the free-streaming evolution, where no relaxation to equilibrium occurs. The pressure asymmetry is even more pronounced at the edge of the fireball, where the density is smaller and the relaxation time longer.
For all of these three calculations, with very different pressure asymmetries,
the resulting transverse flow is almost indistinguishable (Fig. \ref{fig:ptpl}, right panel). It is striking that the transverse flow generated in a kinetic equation involving a relaxation to equilibrium is the same as in a free-streaming evolution. For short evolution times this lack of sensitivity to the pressure asymmetry of the transverse flow can be explained by the flow universality \cite{Vredevoogd:2008id}.
 
\begin{figure}[ht]
\centering
\includegraphics[width=6.2cm,clip]{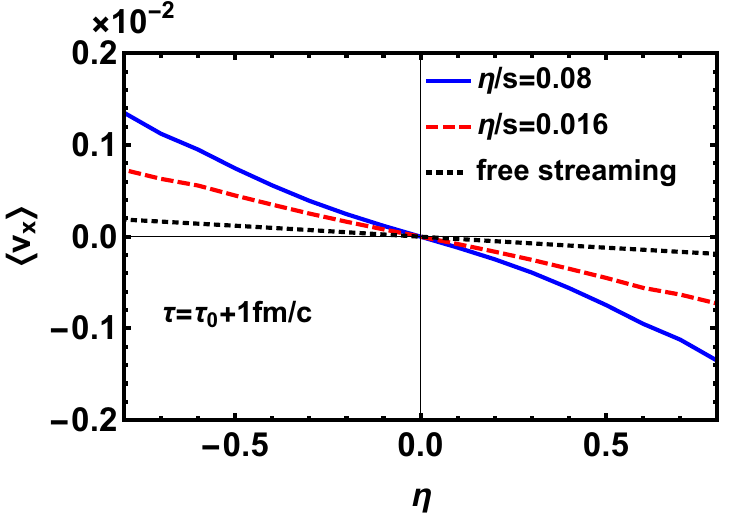}~~~~~\includegraphics[width=6.2cm,clip]{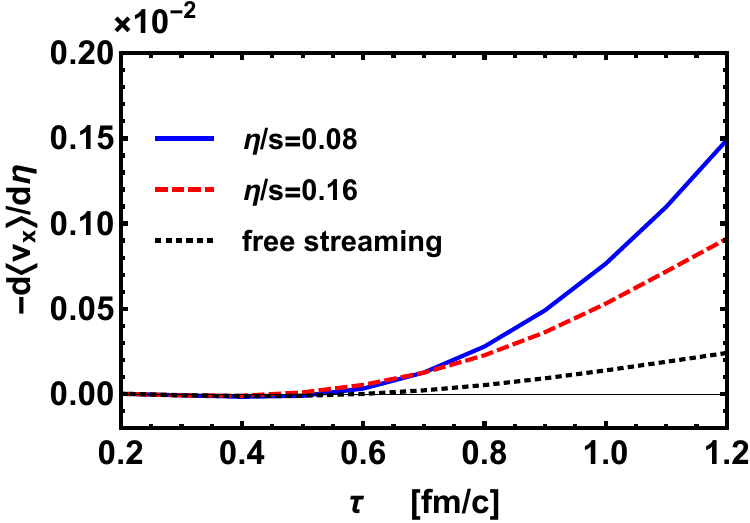}
\caption{(Left panel) The space-time rapidity dependence of the average flow in the transverse direction $x$ after $1$fm/c of kinetic evolution. (Right panel) The time dependence of the slope of the space-time rapidity dependence of the average  flow in the transverse direction $x$. Figs. are from Ref. \cite{Bozek:2022cjj}. }
\label{fig:vx}      
\end{figure}

In the hydrodynamic expansion, the formation of the directed flow from a tilted fireball \cite{Bozek:2010bi}
involves a simultaneous expansion in the longitudinal and transverse directions of the elements of the fluid. The generation of the directed flow results from an  interplay of the transverse and longitudinal pressures in different parts of the fireball. In the kinetic theory calculation it results in a strong sensitivity to differences between the transverse and longitudinal pressures throughout the evolution. The expansion of the tilted fireball in the kinetic theory generates a rapidity odd directed flow (Fig. \ref{fig:vx}, left panel). The average flow in the direction $x$ depends on the space-time rapidity. This mechanism requires a breaking of the forward-backward symmetry due to the tilt of the fireball. It is interesting to note that the resulting directed flow is very sensitive to the relaxation rate. The directed flow is very small for the free-streaming evolution and increases with increasing relaxation rate (decreasing relaxation time).
The space-time rapidity dependence of the directed flow is almost linear and can be parameterized through the slope of this dependence. The slope of this rapidity dependence increases with the evolution time (Fig. \ref{fig:vx}, right panel).
We notice a  very different rate at which the directed flow is generated depending on the relaxation time in the kinetic dynamics.

We  calculate the directed flow generated in a kinetic equation dynamics of the fireball. This requires the solution of the kinetic equation in a system without boost-invariance. Unlike for the transverse flow, we find a  very strong dependence of the directed flow on the pressure asymmetry in the system. This shows that the directed flow could be used as an observable sensitive to nonequilibrium effects at the very early stage of a relativistic nuclear collision. 

\bigskip

{\bf Acknowledgments:} This research is partly supported by
the National Science Centre Grant No. 2018/29/B/ST2/00244.
\bibliography{../../hydr}

\begin{thebibliography}{13}

\bibitem{Heinz:2013th}
U.~Heinz, R.~Snellings, Ann.Rev.Nucl.Part.Sci. \textbf{63}, 123 (2013),
  \texttt{1301.2826}

\bibitem{Gale:2013da}
C.~Gale, S.~Jeon, B.~Schenke, Int.J.Mod.Phys. \textbf{A28}, 1340011 (2013),
  \texttt{1301.5893}

\bibitem{Niemi:2014wta}
H.~Niemi, G.S. Denicol (2014), \texttt{1404.7327}

\bibitem{Florkowski:2013lya}
W.~Florkowski, R.~Ryblewski, M.~Strickland, Phys. Rev. C \textbf{88}, 024903
  (2013), \texttt{1305.7234}

\bibitem{Vredevoogd:2008id}
J.~Vredevoogd, S.~Pratt, Phys. Rev. \textbf{C79}, 044915 (2009),
  \texttt{0810.4325}

\bibitem{Bozek:2022cjj}
P.~Bo{\.z}ek, Phys. Rev. C \textbf{107}, 034916 (2023), \texttt{2212.06018}

\bibitem{Broniowski:2008qk}
W.~Broniowski, W.~Florkowski, M.~Chojnacki, A.~Kisiel, Phys. Rev. \textbf{C80},
  034902 (2009), \texttt{0812.3393}

\bibitem{Kurkela:2018vqr}
A.~Kurkela, A.~Mazeliauskas, J.F. Paquet, S.~Schlichting, D.~Teaney, Phys. Rev.
  C \textbf{99}, 034910 (2019), \texttt{1805.00961}

\bibitem{daSilva:2022xwu}
T.N. da~Silva, D.D. Chinellato, A.V. Giannini, M.N. Ferreira, G.S. Denicol,
  M.~Hippert, M.~Luzum, J.~Noronha, J.~Takahashi (2022), \texttt{2211.10561}

\bibitem{Ambrus:2022koq}
V.E. Ambrus, S.~Schlichting, C.~Werthmann, Phys. Rev. D \textbf{107}, 094013
  (2023), \texttt{2211.14379}

\bibitem{Liyanage:2022nua}
D.~Liyanage, D.~Everett, C.~Chattopadhyay, U.~Heinz, Phys. Rev. C \textbf{105},
  064908 (2022), \texttt{2205.00964}

\bibitem{Kurkela:2018ygx}
A.~Kurkela, U.A. Wiedemann, B.~Wu, Phys. Lett. B \textbf{783}, 274 (2018),
  \texttt{1803.02072}

\bibitem{Bozek:2010bi}
P.~Bo\.zek, I.~Wyskiel, Phys. Rev. \textbf{C81}, 054902 (2010),
  \texttt{1002.4999}

\end{thebibliography}

\end{document}